\documentclass[journal]{IEEEtran}

\makeatletter
\long\def\@makecaption#1#2{\ifx\@captype\@IEEEtablestring%
\footnotesize\begin{center}{\normalfont\footnotesize #1}\\
{\normalfont\footnotesize\scshape #2}\end{center}%
\@IEEEtablecaptionsepspace
\else
\@IEEEfigurecaptionsepspace
\setbox\@tempboxa\hbox{\normalfont\footnotesize {#1.}~~ #2}%
\ifdim \wd\@tempboxa >\hsize%
\setbox\@tempboxa\hbox{\normalfont\footnotesize {#1.}~~ }%
\parbox[t]{\hsize}{\normalfont\footnotesize \noindent\unhbox\@tempboxa#2}%
\else
\hbox to\hsize{\normalfont\footnotesize\hfil\box\@tempboxa\hfil}\fi\fi}
\makeatother
\ifCLASSINFOpdf
  \usepackage[pdftex]{graphicx}
  \graphicspath{{../pdf/}{../jpeg/}}
  \DeclareGraphicsExtensions{.pdf,.jpeg,.png}
\else
  \usepackage[dvips]{graphicx}
  \graphicspath{{../eps/}}
  \DeclareGraphicsExtensions{.eps}
\fi
\usepackage{amsmath}

\usepackage{mathtools}

\usepackage{multicol}
\usepackage{multirow}
\usepackage{url}

\usepackage{algorithm}
\usepackage{algpseudocode}
\usepackage{setspace}

\usepackage{amssymb}
\makeatletter

\newcommand{\Rmnum}[1]{\expandafter\@slowromancap\romannumeral #1@}
\makeatother

\usepackage{arydshln} 
\usepackage[caption=false,font=normalsize,labelfont=sf,textfont=sf]{subfig}
\begin{document}
\title{Distortion-Aware Phase Retrieval Receiver for High-Order QAM Transmission with Carrierless Intensity-Only Measurements}

\author{
    Hanzi~Huang,
    Haoshuo~Chen,
    Qi~Gao,
    Yetian~Huang,
    Nicolas~K.~Fontaine,\\
    Mikael~Mazur,
    Lauren~Dallachiesa,
    Roland~Ryf,
    Zhengxuan~Li,
    and Yingxiong~Song
    
    %\thanks{Manuscript received MON. DAY, YEAR; revised MON. DAY, YEAR.}    
    \thanks{This work was supported in part by the National Key Research and Development Program of China (2021YFB2900801); Science and Technology Commission of Shanghai Municipality (22511100902, 22511100502); 111 project (D20031).
    \textit{(Corresponding author: Haoshuo Chen; Yingxiong Song.)}}% <-this % stops a space
    \thanks{Hanzi Huang, Qi Gao, Yetian Huang, Zhengxuan Li and Yingxiong Song are with Key Laboratory of Specialty Fiber Optics and Optical Access Networks, Shanghai University, 200444 Shanghai, China. (e-mail: hanzihuang@shu.edu.cn; gaoqi21225136@shu.edu.cn; HuangYT@shu.edu.cn; zhengxuanli@shu.edu.cn; herosf@shu.edu.cn).}
    \thanks{Haoshuo Chen, Nicolas K. Fontaine, Mikael Mazur, Lauren Dallachiesa and Roland Ryf are with Nokia Bell Labs, 600 Mountain Avenue, Murray Hill, NJ 07974, USA. (e-mail: haoshuo.chen@nokia-bell-labs.com; nicolas.fontaine@nokia-bell-labs.com; mikael.mazur@nokia-bell-labs.com; lauren.dallachiesa@nokia-bell-labs.com; roland.ryf@nokia.com).}% <-this % stops a space
}

\markboth{}%
{Hanzi \MakeLowercase{\textit{et al.}}: \MakeUppercase{Distortion-Aware Phase Retrieval Receiver for High-Order QAM Transmission with Carrierless Intensity-Only Measurements}}

\maketitle

\begin{abstract}
We experimentally investigate transmitting high-order quadrature amplitude modulation (QAM) signals with carrierless and intensity-only measurements with phase retrieval (PR) receiving techniques.
The intensity errors during measurement, including noise and distortions, are found to be a limiting factor for the precise convergence of the PR algorithm.
To improve the PR reconstruction accuracy, we propose a distortion-aware PR scheme comprising both training and reconstruction stages.
By estimating and emulating the distortion caused by various channel impairments, the proposed scheme enables enhanced agreement between the estimated and measured amplitudes throughout the PR iteration, thus resulting in improved reconstruction performance to support high-order QAM transmission.
With the aid of proposed techniques, we experimentally demonstrate 50-GBaud 16QAM and 32QAM signals transmitting through a standard single-mode optical fiber (SSMF) span of 40 and 80~km, and achieve bit
error rates (BERs) below the 6.25\% hard decision (HD)-forward error correction (FEC) and 25\% soft decision (SD)-FEC thresholds for the two modulation formats, respectively.
By tuning the pilot symbol ratio and applying concatenated coding, we also demonstrate that a post-FEC data rate of up to 140~Gb/s can be achieved for both distances at an optimal pilot symbol ratio of 20\%.
\end{abstract}


% \fi
\IEEEpeerreviewmaketitle

%-------------------------Section 1 ---------------------------%
\section{Introduction}

\IEEEPARstart{C}{oherent} optical communication technology, originally developed for long-haul transmission systems, is now expanding into short-reach transmission applications, including metro, access, and data-center networks, to accommodate the burgeoning traffic growth in edge networks.
Coherent receivers, by interfering the received optical field with a local oscillator (LO) light and directly detecting the signal-carrier beat component, yield a linear representation of the entire optical field as electrical waveforms. 
This naturally facilitates the detection of both amplitude and phase~\cite{fundamentals_of_coherent}.
Transmission impairments like chromatic dispersion (CD), polarization mode dispersion (PMD), and transceiver frequency response impairments can be adaptively compensated via digital signal processing (DSP).
Such capability enables the detection of advanced modulation formats, cementing coherent detection as a particularly compelling approach for a high spectral efficiency~\cite{beyond_1Tbs_Intra-DataCenter,coherent_transmission}.
Yet, the prerequisite of costly, high-stability narrow linewidth lasers prevents coherent detection from being the ideal choice for cost-sensitive short-reach interconnect scenarios.
It also complicates network management as wavelength alignment is required between the transmitters and the receivers.
Consequently, researchers are actively pursuing detection paradigms that meld the merits of both coherent and intensity modulation with direct detection (IM-DD) systems.

\begin{figure}[t!]
	%\vspace{-12pt}
	\centering
	\captionsetup{justification=centering}
    \includegraphics[width=0.88\hsize]{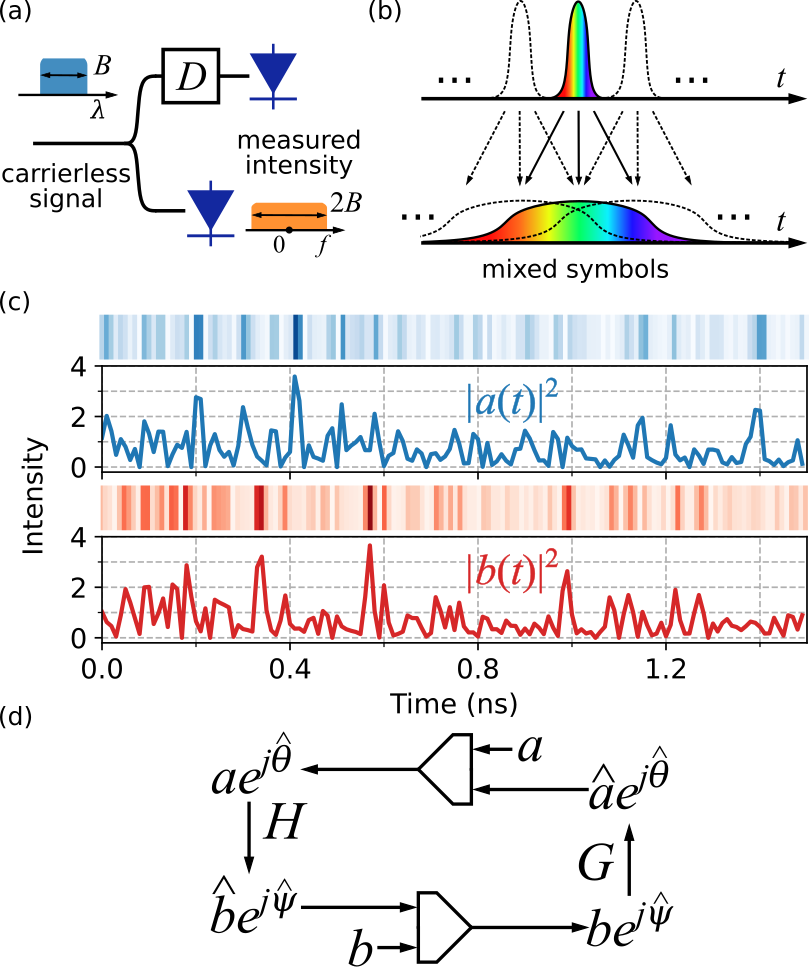}
    
	%\setlength{\abovecaptionskip}{-0.2cm}
	%\setlength{\belowcaptionskip}{-0.5cm}
	%\vspace{-3pt}
	%\vspace{-15pt}
	\centering\caption{Schematics of (a) receiving a carrierless signal by a two-PD-based PR receiver and (b) mixing symbols by chromatic dispersion. (c) Received intensity waveforms on both branches of the PR Rx. (d) Schematic of a GS-algorithm-based reconstruction iteration.}
	\label{fig1}
\end{figure}

To avoid using a LO at the receiver (Rx), the optical carrier is generated inside the transmitter and co-travels with the modulated signal.
This configuration gives rise to various carrier-assisted detection schemes, such as the Kramers-Kronig Rx~\cite{KK,KK_2}, Stokes vector Rx~\cite{Stokes}, and carrier-assisted differential detection~\cite{CADD}.
The concept of carrier-assisted detection schemes is to reconstruct the optical field by harnessing the beat component between the signal and its associated carrier.
Therefore, these schemes necessitate a certain level of carrier-to-signal power ratio (CSPR) to suppress the interference from signal-signal beating, which limits the transmitted signal power, since a substantial fraction of the overall transmitted power is allocated for the carrier.
%In addition, the physical structure of the transmitter demands alteration to adopt optical carriers, usually by coupling the carrier at a suitable power level into the fiber after the modulator, which introduces extra complexities in the design and development of a new Tx hardware structure.
To eliminate the need for any form of optical carrier, various detection schemes utilizing intensity-only (phase-less) and carrierless measurements have been proposed~\cite{MDM_PR,Haoshuo_PR,2D_PD_PR}, termed as phase retrieval (PR).
In the absence of an optical carrier, signal phase information is derived indirectly from the signal-signal beat components generated during photoelectric detection. 
Figure~\ref{fig1}(a) shows the schematic of a two-photodiode(PD)-based PR Rx, which maps the optical signal with an optical bandwidth of $B$ into two radio-frequency waveforms representing measured intensity with an electrical bandwidth of $2B$.
$D$ denotes the dispersion value originating from the dispersive element.
Since the symmetric nature of the signal constellation poses serious phase ambiguity problems, PR becomes an ill-posed problem in optical communication scenarios~\cite{convolutional_PR}.

To solve the PR problem, a strong symbol mixing effect, usually from fiber chromatic dispersion, is preferred to transfer the symbols' phase changes into fluctuations in the received intensity, thereby furnishing vital details about the signal phase, as shown in Fig.~\ref{fig1}(b).
Figure~\ref{fig1}(c) presents the measured intensity fluctuations of a 50~GBaud 16-ary quadrature amplitude modulation (QAM) signal transmitting over a span of 80-km standard single-mode optical fiber (SSMF) and received by a two-PD-based PR Rx.
After symbol mixing, a strong intensity peak implies the overlap symbols modulate similar phases, leading to constructive interference, while vice versa leads to destructive interference, thus leading to a barcode-like pattern in intensity distribution along the time axis.
The reconstruction of the signal needs to reproduce the actual field-based interference models.
In order to achieve this, multiple measurements are induced to allow the algorithm to iterate between different field projection planes.
To mitigate the ill-posedness of the PR problems, prior knowledge about the signal is also induced, such as pilot symbols and spectral constraints, which avoid the algorithm to stagnate in local minima.
By implementing these strategies, a polarization-multiplexed 30~GBaud quadrature phase-shift keying (QPSK) signal is successfully recovered from 4 intensity measurements after a 520-km SSMF transmission based on a modified Gerchberg-Saxton (GS) algorithm~\cite{Haoshuo_PR}.
In the study presented in~\cite{2D_PD_PR}, the authors utilize a two-dimensional photodetector array to expand the intensity observation number to 32, demonstrating that PR signal reconstruction can be achieved without the help of pilot symbols, provided with abundant intensity measurements.
The research in~\cite{weak_carrier} indicates that by digitally incorporating a weak carrier into the signal spectrum to break the inherent symmetry of the optical field, a PR Rx can enhance its performance and save the use of pilot symbols.
Despite it uses an unaltered transmitter structure due to a low CSPR of $\sim$-2~dB, the expansion capability to detect over 2 spatial channels with this scheme can become challenging.
These arise chiefly because the received fields containing inter-channel crosstalk may introduce undesirable interference due to the spatial channel mixing effect, and it is difficult to keep the power of carrier-signal beat components evenly distributed over all spatial channels~\cite{MIMO_interference_canceallation}.
It is further illustrated that a specially designed approach, involving the placement of two carriers at opposite ends of the signal spectrum with orthogonal polarization states, is required to avoid power fading in the received intensities for enabling polarization multiplexing in weak-carrier PR receiving configurations~\cite{pol_mux_weak_PR}.
Conversely, carrierless PR schemes intrinsically support polarization-~\cite{Haoshuo_PR,2D_PD_PR} and mode-division multiplexing transmission~\cite{MDM_PR,MDM_PR_Haoshuo} by augmenting the photodetector count, since the optical power can be evenly distributed over all spatial channels.
The differences distinguishing these approaches suggest that optical carriers, supplementary observations, and pilot symbols possess analogous functions in providing useful information for field reconstruction.
Given the strengths and limitations intrinsic to each method, a practical PR coherent receiver needs to discern and harness the optimal mix of these elements to maximize performance.

One effective method for enhancing the data rate of PR receivers is utilizing advanced modulation formats.
Such an approach not only leads to higher spectral efficiency but also optimally exploits the reconstruction accuracy offered by each converged iteration cycle.
To accommodate high-order QAM transmission, recently, several works have focused on improving the convergence speed and accuracy of the PR Rx by driving the GS algorithm out of stagnation more aggressively based on adaptive intensity transformation~\cite{AIT_PR} and weight decision constraint operations~\cite{WD_PR}, which show distinct convergence speed improvements in both simulations and experiments.
Whereas, these methods show mild improvements in the converged bit error rate (BER) levels in experiments, which cannot drive the 25-GBuad 16QAM signal below 7\% forward error correction (FEC) BER threshold, implying that the reconstruction integrity issue of PR Rx to support advanced modulation formats in the presence of complicated channel impairments is not fully addressed.
%Although the feasibility of PR field reconstruction has been experimentally demonstrated, its reconstruction integrity to support advanced modulation formats in the presence of complicated channel impairments is not fully addressed.
Transmission impairment compensation techniques in standard coherent DSP procedures cannot directly apply to PR schemes.
This is because the fundamental channel linearization assumption ceases to be valid during the square-law detection process.
%, which eliminates signal phase information.
Furthermore, cascading standard coherent DSP procedures after field reconstruction could lead to sub-optimal algorithm convergence as the signal distortion during transmission is not fully considered by the algorithm, thus resulting in performance degradation.

This paper aims to fill the existing blank in the channel impairment estimation and compensation techniques within PR receiving schemes.
In this paper, we study the limiting factor of PR convergence under noisy measurements, and experimentally investigate the reconstruction performance boundary of a two-PD-based PR receiver by exploring its feasibility in transmitting 50~GBaud 16QAM and 32QAM signals over one SSMF span, given that sufficient symbol mixing and prior knowledge from pilot symbols are provided.
In this case, as the intrinsic signal recovery capability of the algorithm is ensured, the main limiting factors of the final performance are the errors in the detected intensity waveforms, which do not obey the propagation rule described by the dispersion transfer equation in the GS iteration.
To mitigate the intensity errors stemming from signal distortions, we propose a distortion-aware PR scheme comprising both training and reconstruction stages.
%Within the training stage, various channel impairments, such as Tx and Rx frequency response, IQ-dependent impairments, and nonlinear transfer effect of the modulator, are estimated by minimizing errors between the propagated training sequences' intensities and the actual measurements.
%During the reconstruction stage, the proposed PR scheme empowers the algorithm to emulate distortions caused by various channel impairments throughout its iterations.
By estimating and emulating the distortion caused by various channel impairments throughout the PR DSP, the proposed strategy fosters enhanced congruence between the estimated and measured amplitude during iterations, thus enabling improved reconstruction performance to support high-order QAM transmission.
With the aid of the proposed techniques, for both 40- and 80-km SSMF transmission, the BERs of recovered 16QAM and 32QAM signals can reach under below the 6.25\% hard decision (HD)-FEC threshold of $4.7\times10^{-3}$~\cite{6p25_HD_FEC} and 25\% soft decision (SD)-FEC threshold of $4\times10^{-2}$~\cite{25_SD_FEC_2}, respectively.
Additionally, by tuning the pilot symbol ratio, we demonstrate that our proposed PR scheme can support a post-FEC data rate of up to 140~Gb/s per polarization for a single wavelength channel for both distances at an optimal pilot symbol ratio of 20\%.

For the rest of the paper, Section \Rmnum{2} delineates the principle and methodology of the proposed PR scheme.
Section \Rmnum{3} outlines the experimental setup for proof-of-concept validation.
Section \Rmnum{4} presents the experimental results from the aspects of channel estimation, field reconstruction, and achievable data rate.
Section \Rmnum{5} concludes the paper.

%For the notations, regular lowercase or uppercase letters denote scalars.
%Boldface lowercase and uppercase letters denote vectors and matrices, respectively.
%%We use $(.)^T$, $(.)^*$ and $(.)^H$ to denote transpose, conjugate and transpose conjugation operation.
%$\mathbb{R}$ and $\mathbb{C}$ denotes the set of real and complex number, respectively.
%$\mathbb{E}\left[.\right]$ denotes the mathematical expectation.% and $\Vert.\Vert^2$ denotes the squared Euclidean norm.
%%$\boldsymbol{I}_{N}$ denotes identity matrix with size of $N$.
%%$\boldsymbol{0}_{N}$ and $\boldsymbol{0}_{N,M}$ denote zero matrices with sizes of $N\times N$ and $N\times M$.

%-------------------------Section 2 ---------------------------%
\section{Principle and Methodology}

\subsection{Phase Retrieval Convergence Limit}

The convergence property of the GS algorithm has been proved mathematically when it is proposed in \cite{GS_algorithm}, while its convergence limit, which decides the ultimate convergence accuracy of the GS-based PR receiving scheme, is not clear.
In this subsection, we aim to analyze the convergence limit of GS algorithm under noisy amplitude measurements by derivations.
First, we consider a standard GS-based PR iteration with two alternated projections, with its schematic shown in Fig.~\ref{fig1}(d).
The goal is to estimate the field phase distribution $\boldsymbol{\theta} \in \mathbb{R}^{N \times 1}$ and $\boldsymbol{\psi} \in \mathbb{R}^{N \times 1}$ with knowing the measured amplitude distribution $\boldsymbol{a} \in \mathbb{R}^{N \times 1}$ and $\boldsymbol{b} \in \mathbb{R}^{N \times 1}$ across $N$ time samples.
The relationship of the $i$-th sample of the fields at two projection planes is given by
\begin{equation}
\boldsymbol{b}_i e^{j\boldsymbol{\psi}_i}=\sum_{k=1}^{N}\boldsymbol{H}_{i,k}\boldsymbol{a}_k e^{j\boldsymbol{\theta}_k}
\end{equation}
and
\begin{equation}
\boldsymbol{a}_i e^{j\boldsymbol{\theta}_i}=\sum_{k=1}^{N}\boldsymbol{G}_{i,k}\boldsymbol{b}_k e^{j\boldsymbol{\psi}_k},
\end{equation}
where $i=1,...,N$, $\boldsymbol{H} \in \mathbb{C}^{N \times N}$ and $\boldsymbol{G} \in \mathbb{C}^{N \times N}$ denote the Toeplitz matrix derived by shifting the channel impulse response.
We assume that $\boldsymbol{H}$ and $\boldsymbol{G}$ are known beforehand. $\hat{\boldsymbol{\theta}}$ and $\hat{\boldsymbol{\psi}}$ denote the estimated phase during the iteration.
By propagating the estimated field $\boldsymbol{a}e^{j\hat{\boldsymbol{\theta}}}$ to the other projection plane, the expected amplitude of the $i$-th sample of estimated field $\boldsymbol{b}e^{j\hat{\boldsymbol{\psi}}}$ is
\begin{equation}
\hat{\boldsymbol{b}}_i=\left|\sum_{k=1}^{N}\boldsymbol{H}_{i,k}\boldsymbol{a}_k e^{j\hat{\boldsymbol{\theta}}_k}\right|.
\end{equation}
The gradient vector of combining measured amplitude with estimated phase is given by 
\begin{equation}
\nabla \boldsymbol{f}(\hat{\boldsymbol{b}},\hat{\boldsymbol{\psi}})=\boldsymbol{b}e^{j\hat{\boldsymbol{\psi}}}-\hat{\boldsymbol{b}}e^{j\hat{\boldsymbol{\psi}}}
\end{equation}
If we assume the estimated phase $\hat{\boldsymbol{\theta}}$ and $\hat{\boldsymbol{\psi}}$ approximate their true values, the mathematical expectation of the amplitude squared error between the measured and estimated value would approach zero as
\begin{equation}
\mathbb{E}\left[\left(\boldsymbol{b}_i-\hat{\boldsymbol{b}}_i\right)^2\right]\rightarrow 0, \text{as}\ \hat{\boldsymbol{\theta}}\rightarrow\boldsymbol{\theta}\ \text{and}\ \hat{\boldsymbol{\psi}}\rightarrow\boldsymbol{\psi}.
\label{amplitude_squared_error}
\end{equation}
Equation~(\ref{amplitude_squared_error}) shows that each element in $\nabla \boldsymbol{f}(\hat{\boldsymbol{b}},\hat{\boldsymbol{\psi}})$ and $\nabla \boldsymbol{f}(\hat{\boldsymbol{a}},\hat{\boldsymbol{\theta}})$ has a tendency to approach zero, implying that under no amplitude measurement error, the correct phase $\boldsymbol{\theta}$ and $\boldsymbol{\psi}$ become a stationary point of the algorithm.

Now we consider that the amplitudes contain zero-mean additive white Gaussian noise (AWGN) during observation given by $\widetilde{\boldsymbol{a}}=\boldsymbol{a}+\boldsymbol{n}$ and $\widetilde{\boldsymbol{b}}=\boldsymbol{b}+\boldsymbol{w}$,
%\begin{equation}
%\widetilde{\boldsymbol{a}}=\boldsymbol{a}+\boldsymbol{n},\ %\widetilde{\boldsymbol{b}}=\boldsymbol{b}+\boldsymbol{w},
%\label{amplitude_w_noise}
%\end{equation}
where $\boldsymbol{n}$ and $\boldsymbol{w}$ represent independent random noise.
%Using these definitions in Eq.~(\ref{amplitude_w_noise}), 
Using these definitions, Eq.~(\ref{amplitude_squared_error}) can be rewritten as
\begin{equation}
\begin{aligned}
\mathbb{E}\left[\left(\widetilde{\boldsymbol{b}}_i-\hat{\boldsymbol{b}}_i\right)^2\right]=&\mathbb{E}\left[\left(\widetilde{\boldsymbol{b}}_i-\left|\sum_{k=1}^{N}\boldsymbol{H}_{i,k}\widetilde{\boldsymbol{a}}_k e^{j\hat{\boldsymbol{\theta}}_k}\right|\right)^2\right]
    \\=&\mathbb{E}\left[\boldsymbol{w}^2+\left|\sum_{k=1}^{N}\boldsymbol{H}_{i,k} \boldsymbol{n} e^{j\hat{\boldsymbol{\theta}}_k} \right|^2\right]
    \\\rightarrow&W_0^2+N_0^2,\ \text{as}\ \hat{\boldsymbol{\theta}}\rightarrow\boldsymbol{\theta}\ \text{and}\ \hat{\boldsymbol{\psi}}\rightarrow\boldsymbol{\psi},
\end{aligned}
\label{amplitude_squared_error_w_noise}
\end{equation}
where $W_0^2$ and $N_0^2$ denotes the noise variance of $\boldsymbol{w}$ and $\boldsymbol{n}$.
The second step in Eq.~(\ref{amplitude_squared_error_w_noise}) is derived by expanding the square bracket and simplifying the expression by $\mathbb{E}\left[\boldsymbol{n}\right]=0$ and $\mathbb{E}\left[\boldsymbol{w}\right]=0$.
The third step assumes a precondition as $\mathbb{E}\left[\sum_{k=1}^{N}|\boldsymbol{H}_{i,k}|^2\right]=1$, since the propagation operation obeys energy conservation.
Compared with Eq.~(\ref{amplitude_squared_error}), the discrepancy in Eq.~(\ref{amplitude_squared_error_w_noise}) shows that, whether the convergence capability of an alternative projection PR scheme is good or not, its achievable performance limit is fundamentally constrained by the noise.
As the observation error limits the minimum stepsize of the gradients during iterations, the estimated field will be forced to wander around the correct values by the misleading gradients, which limits the convergence accuracy.%, as shown in \underline{Fig.~()}.

To achieve a more accurate approximation between the estimated and measured amplitudes, the power of $\boldsymbol{n}$ and $\boldsymbol{w}$ should be suppressed. 
While it is impossible to predict amplitude errors stemming from noise due to its nature of randomness, a portion of these errors arises due to signal distortion, which comes from the non-ideal transfer characteristic of the devices. % and is usually invariant with time.
Therefore, recognizing the precise form of the distortion could partially compensate its influences and lead to reconstruction results with higher accuracy.

\begin{figure*}[htbp!]
	%\vspace{-12pt}
	\centering
	\captionsetup{justification=centering}
    \includegraphics[width=0.92\hsize]{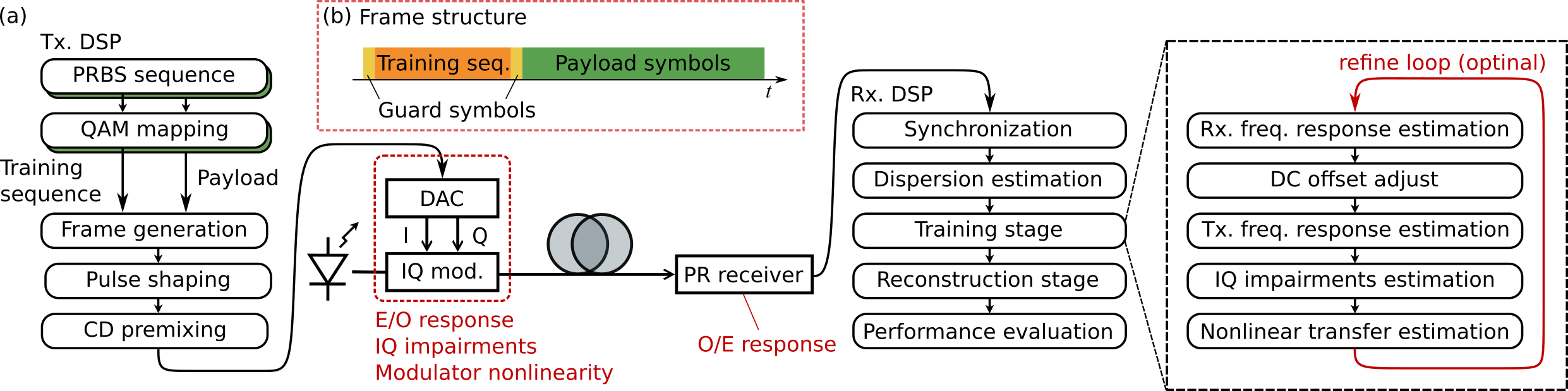}
    
	%\setlength{\abovecaptionskip}{-0.2cm}
	%\setlength{\belowcaptionskip}{-0.5cm}
	%\vspace{-3pt}
	%\vspace{-15pt}
	\centering\caption{(a) Schematics of the distortion-aware PR system with the transceiver DSP flows and (b) its frame structure.}
	\label{fig2}
\end{figure*}

\subsection{Distortion-aware Phase Retrieval}

Figure~\ref{fig2}(a) illustrates the schematic of the PR transmission system as well as the transceiver DSP procedures employed in the experiment.
Upon completing QAM mapping, frame generation, and pulse shaping, the transmitter (Tx) DSP premixes the sequence with extra chromatic dispersion in the digital domain, ensuring pronounced intensity fluctuations on the Rx end.
The premixed patterns undergo conversion via the digital to analog converter (DAC), subsequently modulated into the optical domain, and then transmitted via fiber to be captured by a PR receiver. 
The receiver is based on a two-PD-based PR configuration with the same structure as in Fig.~\ref{fig1}(a).
Throughout the transmission, several impairment factors, such as the electrical-to-optical (E/O) and optical-to-electrical (O/E) frequency response, in-phase and quadrature (IQ)-dependent impairments, and modulator nonlinearity, may distort the signals.

To compensate for the distortion, a distortion-aware PR scheme is proposed, which uses a training sequence to identify the channel impairments.
The frame structure is depicted in Fig.~\ref{fig2}(b).
The training sequence is placed at the start of a frame, with two segments of guard symbols placed beside the training sequence to prevent unknown mixing symbols from the payload section after chromatic dispersion.
These guard symbol sequences can be derived as the cyclic prefix and postfix of the training sequence to make it follow the circular convolution.
The pilot symbols distribute across the payload section evenly to facilitate convergence by eliminating phase ambiguity.
The Rx DSP employs the pre-known training sequence to conduct dispersion estimation, time synchronization, and determine channel impairments during the training stage.
The dispersion values for the two branches are estimated using a grid-search method aimed at maximizing the correlation peak values between the propagated training sequence's intensities and received intensities.
Within the training stage, the channel impairments are estimated by minimizing errors between the distorted forward propagated training sequences' intensities and the actual measurements.
During the reconstruction stage, the proposed PR scheme empowers the algorithm to emulate distortions caused by channel impairments in forward propagation and apply reverse distortion in backward propagation throughout its iterations.
Besides these procedures, the receiver DSP also encompasses a performance evaluation module to calculate metrics like BERs, generalized mutual information (GMI), and the recovered constellations.

\subsection{Training Stage}
The whole training stage contains several procedures, including estimating Tx and Rx response, optimal direct current (DC) level, IQ-dependent impairments (including power imbalance, time skew, and phase mismatch), and modulator nonlinearity.

To start with, the Rx response estimation module is for obtaining the channel distortion induced by the frequency roll-off during the O/E process, e.g. bandwidth limitations raised from the photodiodes and high-speed radio frequency (RF) cables as well as connectors.
As these impairments happen after photoelectric detection, a pair of feed-forward equalizers (FFEs) are used for compensating those linear impairments. 
To calibrate the FFEs, the training sequence is initially propagated to the receiver photodiodes using the estimated chromatic dispersion values.
The expected waveforms of received intensities are then computed via square-law detection.
Subsequently, DC components from both received and expected intensities are removed, and FFE is trained on these DC-removed intensities.
After FFE equalization, the DC offset levels are reintroduced, with their values optimized to minimize the mean absolute error (MAE) between the received and expected intensities.
The aforementioned procedures eliminate the impact of incorrect DC offset level under O/E frequency response, thus providing the PR algorithm an optimal baseline of complete darkness.

For the estimation of Tx response, the situation becomes intricate, given that the I- and Q-channels can exhibit independent impairments, meaning their frequency-dependent phase changes need to be estimated.
Since no direct phase information about the channel is measured, this transforms into a PR problem.
Several works in~\cite{Haoshuo_PR} and~\cite{2D_PD_PR} have successfully implemented complex-valued channel estimation using intensity measurements over multiple photodiodes.
These primarily rely on multiple observations as a fundamental condition to address the PR challenge.
In contrast, this paper introduce a complex-valued channel estimation technique that only employs one single PD measurement, with its pseudocode provided in Algorithm~\ref{alg:tx_est}.
$p(t)$, $h_{CD}(t)$, $h_{BW}(t)$ denote the complex-valued field of training sequence, the impulse response of fiber CD, and the impulse response for a bandwidth constraint filter used to eliminate out-of-band frequency components, respectively. 
The main concept is merging the phases of the propagated training sequence $\hat{p}_{\rm{d,rx}}(t)$ with the measured amplitudes $a(t)$ to derive the estimated distorted field $\tilde{p}_{\rm{d,rx}(t)}$, and estimating the channel with least square (LS) algorithm iteratively until the MAE between $\hat{p}_{\rm{d,rx}}(t)$ and $a(t)$ starts to increase or the iteration number reaches a preset maximum $K$.
The convolution operation is used for conciseness in expression but is preferred to be implemented in frequency domain to achieve fast computation.
In the experiment, we set $K$ as 20 and observe that the estimation converges within 8 iterations in most cases, which can be attributed to the initial $\hat{p}_{\rm{d,rx}}(t)$ is close to its correct values.
By fully leveraging the prior knowledge of the training sequence, the proposed design alleviates the effects of channel disparities across multiple measurements, including bandwidth limitation variations among PDs and additional distortion introduced by dispersive elements.
This could lead to improved accuracy and versatility regardless of the number of PDs.

\begin{algorithm}[htbp]
\setstretch{1.3}
\caption{Transmitter Response Estimator}
	\begin{algorithmic}[1]
    %\Function{Tx\_estimator}{$p(t)$,$a(t)$,$CD$,$K$}
    \State $\hat{h}^{(0)}_{\rm{tx}} \gets \delta(t)$\Comment{Initial guess of Tx. response}
    \For {$k=1,2,\ldots,K$}
        \State $\hat{p}_{\rm{d}}(t) \gets p(t)*\hat{h}^{(k-1)}_{\rm{tx}}(t)$\Comment{Calculate distorted signal}
        %\State \textbf{if in} refine loop \textbf{then} distort $\hat{p}_{\rm{d}}(t)$ considering IQ impairments and nonlinear transfer function
        \State $\hat{p}_{\rm{d,rx}}(t) \gets \hat{p}_{\rm{d}}(t)*h_{CD}(t)$\Comment{Propagate to Rx.}
        \State ${\rm MAE}=\mathbb{E}\big[\left|\hat{p}_{\rm{d,rx}}(t)-a(t)\right|\big]$\Comment{Mean absolute error}
        \State \textbf{if} MAE starts to increase \textbf{then break}
        \State $\tilde{p}_{\rm{d,rx}}(t) \gets a(t)\exp{j\angle{\hat{p}_{\rm{d,rx}}(t)}}$\Comment{Combine amplitude}
        \State $\tilde{p}_{\rm{d}}(t) \gets \tilde{p}_{\rm{d,rx}}(t) * h_{\rm{BW}}(t) * h^{-1}_{CD}(t)$\Comment{Back to Tx.}
        %\State \textbf{if in} refine loop \textbf{then} apply reverse distortion to $\tilde{p}_{\rm{d}}(t)$ 
        \State Calculate $\hat{h}^{(k)}_{\rm{tx}}(t)$ with LS channel estimation using $\tilde{p}_{\rm{d}}(t)$ as received signals and $p(t)$ as transmitted patterns
    \EndFor
    \State \Return $\hat{h}^{(k)}_{\rm{tx}}(t)$
    %\EndFunction
    \end{algorithmic}
    \label{alg:tx_est}
\end{algorithm}

For the estimation of IQ-dependent impairments and modulator nonlinearity, the distortions are introduced by field-based mathematical models.
The distorted optical field due to IQ-dependent impairments can be expressed as
\begin{equation}
s_{\rm IQ}(t)=I(t)+j\sqrt{1+\rho}\cdot Q(t+\tau)e^{j\phi},
\end{equation}
where variables $\rho$, $\tau$, $\phi$ denote the power imbalance coefficient, time skew, and phase error between the I- and Q-channels, respectively. 
$I(t)$ and $Q(t)$ denote the undistorted real-valued I- and Q-channel signals.
When considering modulator nonlinearity, the nonlinear transfer function for the I-channel is approximated using a cubic polynomial, given by
\begin{equation}
I_{\rm NL}(t)=I(t)+c_I^{(2)}I^2(t)+c_I^{(3)}I^3(t).
\end{equation}
The Q-channel follows a similar representation, with nonlinear coefficients denoted as $c_Q^{(2)}$ and $c_Q^{(3)}$.
Note that the modulator nonlinearity is applied before the IQ impairments as two Mach-Zehnder interferometer can have different nonlinear properties, even it is estimated at the last step.
These parameters are determined by minimizing the mean absolute error between the distorted forward propagated training sequence's intensities and the actual measured values (both are power normalized).
To be specific, the optimization employs a grid-search method in a greedy algorithm manner.
The search process starts with a zero value for each parameter, and then a parameter scan is conducted with a suitable search span and granularity sequentially in an order of $\phi$, $\tau$, $\rho$, $c_I^{(2)}$, $c_Q^{(2)}$, $c_I^{(3)}$, $c_Q^{(3)}$ over three rounds.

After completing an entire training cycle, the scheme offers an optional refinement loop to iteratively enhance estimation accuracy, as indicated by the red arrow in Fig.~\ref{fig2}(a).
This is mainly because the Rx frequency limitations can be over estimated in the first cycle, since the frequency response impairments from Tx has not been considered when training the FFEs.
Meanwhile, knowledge about the distortion induced by IQ-dependent impairments and modulator nonlinearity can be employed to modify the third line of Algorithm~\ref{alg:tx_est} to further increase its approximation capability in the refinement loop.
%\begin{algorithm}[htbp]
%\setstretch{1.3}
%\caption{Distortion-aware Phase Retrieval}
%	\begin{algorithmic}[1]
%    %\Function{Tx\_estimator}{$p(t)$,$a(t)$,$CD$,$K$}
%    \State $\hat{s}(t)=\hat{s}_0(t)$\Comment{Initial guess of optical field}
%    \For {$n=1,2,\ldots,N$}
%        \State $\hat{s}(t) \gets \hat{s}(t)*h_{CD\rm{,premix}}(t)$\Comment{Apply CD premixing}
%        \State Apply Tx. distortion to $\hat{s}(t)$
%        \State $\hat{s}(t) \gets \hat{s}(t)*h_{CD\rm{,fiber}}(t)$\Comment{Apply fiber CD}
%        \State $\hat{s}(t) \gets a(t)\exp{j\angle{\hat{s}(t)}}$\Comment{Combine amplitude}
%        \State $\hat{s}(t) \gets \hat{s}(t)*h_{D}(t)$\Comment{Apply dispersion element CD}
%        \State $\hat{s}(t) \gets b(t)\exp{j\angle{\hat{s}(t)}}$\Comment{Combine amplitude}
%        \State $\hat{s}(t) \gets \hat{s}(t)*h_{D+CD,\rm{fiber}}^{-1}(t)$\Comment{Back to Tx.}
%        \State Apply Tx. reverse distortion to $\hat{s}(t)$
%        \State $\hat{s}(t) \gets \hat{s}(t)*h_{CD\rm{,premix}}^{-1}(t)$\Comment{Reverse CD premixing}
%        \State $\hat{s}(t_{\rm{pilot}}) \gets s(t_{\rm{pilot}})$\Comment{Apply pilot symbols}
%        \State $\hat{s}(t_{\rm{odd}}) \gets 0$\Comment{Remove odd samplings}
%        \State $\hat{s}(t) \gets \hat{s}(t)*h_{\rm{RRC}}(t)$\Comment{Apply pulse shaping}
%        \State $\hat{s}(t) \gets \hat{s}(t)/\sqrt{\mathbb{E}\left[|\hat{s}(t)|^2\right]}$\Comment{Power normalization}
%    \EndFor
%    \State \Return $\hat{s}(t)$
%    %\EndFunction
%    \end{algorithmic}
%    \label{alg:htx_est}
%\end{algorithm}

\subsection{Reconstruction Stage}

\begin{figure}[htbp]
	%\vspace{-12pt}
	\centering
	\captionsetup{justification=centering}
    \includegraphics[width=0.99\hsize]{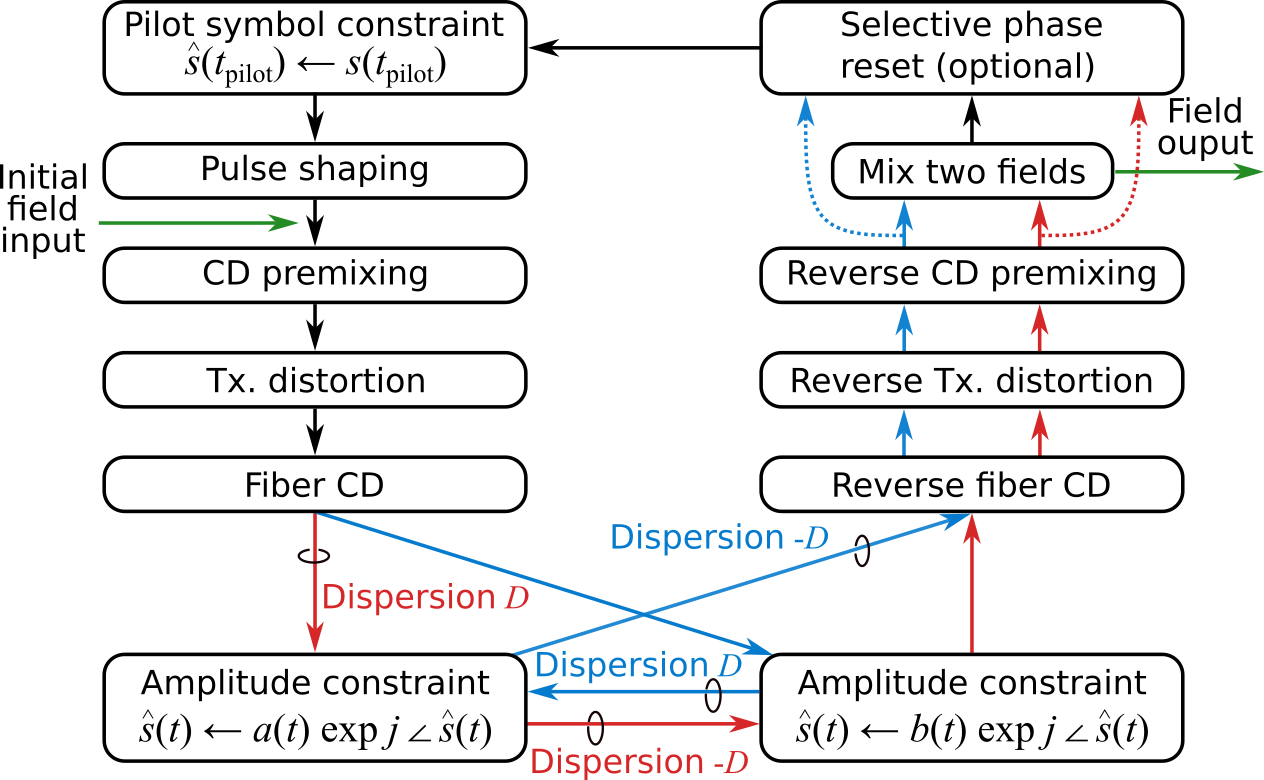}
    
	%\setlength{\abovecaptionskip}{-0.2cm}
	%\setlength{\belowcaptionskip}{-0.5cm}
	%\vspace{-3pt}
	%\vspace{-15pt}
	\centering\caption{Schematic of DSP procedures in the reconstruction stage.}
	\label{fig3}
\end{figure}

The reconstruction stage of the proposed PR scheme is illustrated in Fig.~\ref{fig3},
which resembles the pilot symbol-assisted PR in~\cite{Haoshuo_PR,Haoshuo_phase_reset}, albeit with several modifications.

The first modification is inducing the Tx distortion raised from Tx frequency response, IQ impairments, and modulator nonlinearity when the estimated field is forward propagated from Tx to Rx.
Correspondingly, a designated module introduces reverse distortion during backward propagation of the estimated field from Tx to Rx.
This reverse distortion channel is derived by inversely determining the estimated channel distortion.
The reverse Tx response is determined by inverting the Tx response in the frequency domain.
For the reverse IQ impairment parameters, they are ascertained by taking the negative values of the estimated parameters $\rho$, $\tau$, $\phi$.
The reverse nonlinear transfer function is deduced by computing the inverse function of the estimated cubic polynomial function representing modulator nonlinearity.

The second modification is inducing a dual-trace propagation mechanism, as highlighted in red and blue in Fig.~\ref{fig3}.
The two color arrows denote the two estimated field traces experiencing different amplitude constraint sequences.
After amplitude constraints, these two traces approach the correct field via distinct trajectories, enabling higher accuracy and reducing the likelihood of convergence to local minima after field mixing in the Tx plane.
Moreover, the estimated results of these traces should coincide upon reaching the correct phase estimation.
Consequently, any pronounced discrepancy in their amplitude waveforms indicates potential estimation failures, thus providing reliable criterion for the selective phase reset module~\cite{Haoshuo_phase_reset}.
In the experiment, the selective phase reset provides better reconstruction results under an insufficient pilot symbol ratio, while it shows a mild influence to the reconstruction results at a high symbol ratio such as 50\%.%, which is further discussed in subsection \Rmnum{4}-C.
\begin{figure*}[t!]
	%\vspace{-12pt}
	\centering
	\captionsetup{justification=centering}
    \includegraphics[width=0.8\hsize]{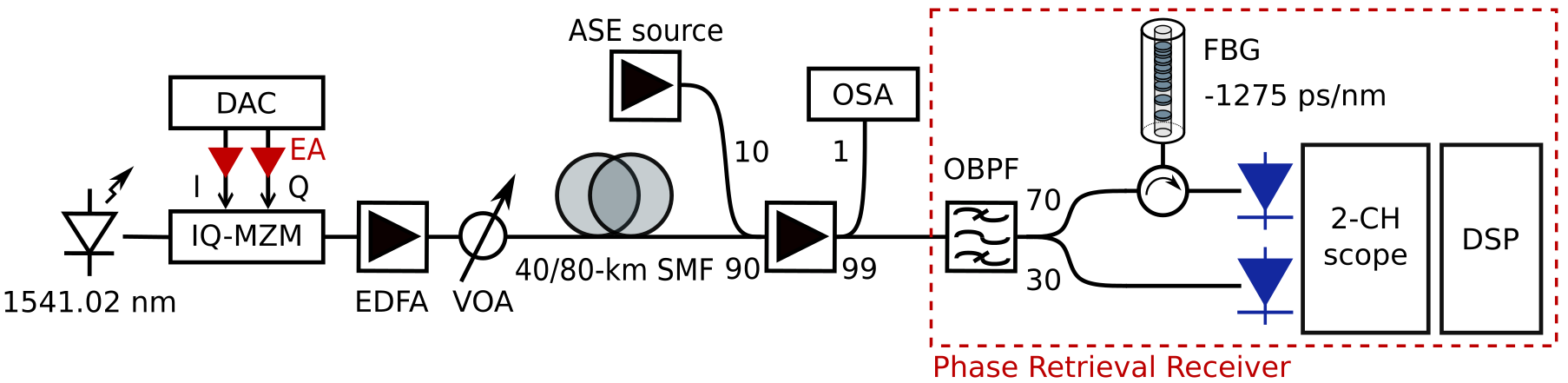}
    
	%\setlength{\abovecaptionskip}{-0.2cm}
	%\setlength{\belowcaptionskip}{-0.5cm}
	%\vspace{-3pt}
	%\vspace{-15pt}
	\centering\caption{Experimental setup for transmission over 40- and 80-km SSMF via the two-PD-based PR receiver.}
	\label{fig4}
\end{figure*}

In addition to the aforementioned modifications, the proposed PR scheme integrates a forward and backward CD premixing module to adapt digitally CD-premixed symbols.
Besides, every CD propagation operation includes a bandwidth constraint filter designed to eliminate local minima with incorrect frequency structures.
After implementing the pilot constraint on the Tx projection plane, the estimated field undergoes down-sampling to the symbol rate, and then a pulse shaping procedure is introduced to circumvent the estimation of sub-symbol sampling points, thereby reducing the degree of freedom for the optimization problem.
Note that the pilot symbol constraint has to be conducted at the transmitter projection plane to guarantee the one-to-one mapping between ideal and estimated pilot symbols as there is no channel memory effect in this projection.

%-------------------------Section 3 ---------------------------%
\section{Experimental Setup}

Figure~\ref{fig4} presents the experimental setup for the proposed PR scheme transmitting 50-GBaud QAM signals over a 40-km or 80-km span of SSMF.
At the transmitter side, a continuous wave (CW) laser, featuring a 10-kHz linewidth (Pure Photonics PPCL500) and operating at 1541.02~nm, is modulated by an IQ-Mach-Zehnder modulator (MZM) to produce the premixed QAM signal with a roll-off factor of 1\%.
The MZM biases are carefully set at the null point to fully suppress the carrier.
The transmitted IQ patterns undergo the Tx DSP procedures as illustrated in Fig.~\ref{fig2}(a).
The payload and training symbol segments are generated by pseudorandom bit patterns independently with no correlation.
The training sequence consists of $2^{13}$ QPSK symbols, positioned between two guard symbol segments, each of which has a length of 64 symbols.
The payload sequence circularly repeats a segment of $2^{13}$ QAM symbols for 30 times.
The digital CD for premixing is set to -3000~ps/nm, which corresponds to overlapping $\sim$60 symbols at 50~GBaud.
This large amount of CD premixing aims to explore the PR Rx performance under extensive symbol mixing.
In practice, as SSMF provides a unit CD of $\sim$17~ps/nm/km, the digital CD premixing can be removed when the system operates at a higher symbol rate or transmits through longer distances.
An extra 0.5\% clipping is applied before the digital patterns are sent to the DAC to constrain the peak-to-average power ratios (PAPRs), thus improving the power efficiency of electric amplifiers (EAs).
The output modulated signal is pre-amplified by an erbium-doped fiber amplifier (EDFA), followed by a variable optical attenuator (VOA), which is for controlling the transmitting optical power to eliminate nonlinear distortion during fiber transmission, with a launch power of -2 and \mbox{-2.5~dBm} for the 40- and 80-km transmission, respectively.

At the receiver side, another EDFA is used to re-boost the received optical power, with its input port connected to an amplified spontaneous emission (ASE) source by a 90:10 optical coupler.
The coupled ASE power can be adjusted by tuning the pump current of the ASE source, resulting in received signals with different optical signal-to-noise ratio (OSNR) values into the PR Rx.
The output port of the Rx EDFA connects to a 99:1 optical splitter, with the 1\% branch connected to an optical spectrum analyzer (OSA) for OSNR measurement.
The OSNR is measured as the ratio of the signal power to the ASE power within 0.1~nm bandwidth.
Inside the PR Rx, an optical band-pass filter (OBPF) is used to filter out-of-band noise with a bandwidth of $\sim$0.6~nm, cascaded by a 70:30 optical splitter.
This power splitting ratio is chosen to balance the received optical power at the two photodiodes.
The Rx EDFA pump current is adjusted to set the filter output power as $\sim$7~dBm when no extra ASE is introduced.
The dispersive element used in the PR Rx is a fiber Bragg grating (FBG)-based dispersion compensation module (Proximion DCM-FB) with a measured chromatic dispersion of -1275~ps/nm at the operating wavelength.
In comparison to using a span of SSMF or dispersion compensating fiber (DCF) as the dispersive element, the FBG-based approach offers several advantages: it introduces mild loss ($\sim$3~dB), negligible latency ($\sim$115~ns, including the delay from patch cord fiber), and zero nonlinear effects.
Additionally, it can provide a distinct amount of dispersion, making it particularly suitable for PR systems that aim at a high reconstruction accuracy.
In practice, as continuous dispersion can be offered over the whole C-band via FBG-based solution, a single dispersive element can be shared by multiple PR receivers operating at different wavelength channels, thus reducing the implementation cost.
The received two-branch optical signals are then detected by two 50-GHz class photodiodes (u$^2$t XPDV2120RA).
The detected electrical signals are then captured by a digital storage oscilloscope with a sampling rate of 160~GSa/s.
The captured waveforms experience down-sampling to 2 samples per symbol, power normalization, and then being processed offline with the Rx DSP flow illustrated in Fig.~\ref{fig2}(a).
The tap number of FFEs for Rx response estimation is set to 101, and the tap number for Tx response estimation in Algorithm~\ref{alg:tx_est} is adjusted to 511.
The relatively high tap number for the latter could be attributed to the distortions and reflections along the RF chain connecting discrete components at the Tx, and it could also be due to that having sufficient tap coefficients helps solving the PR problem by enabling more refined distortion estimation.

%-------------------------Section 4 ---------------------------%
\section{Experimental Results}

\subsection{Channel Estimation Results}

\begin{figure}[htbp]
	%\vspace{-12pt}
	\centering
	\captionsetup{justification=centering}
    \includegraphics[width=0.98\hsize]{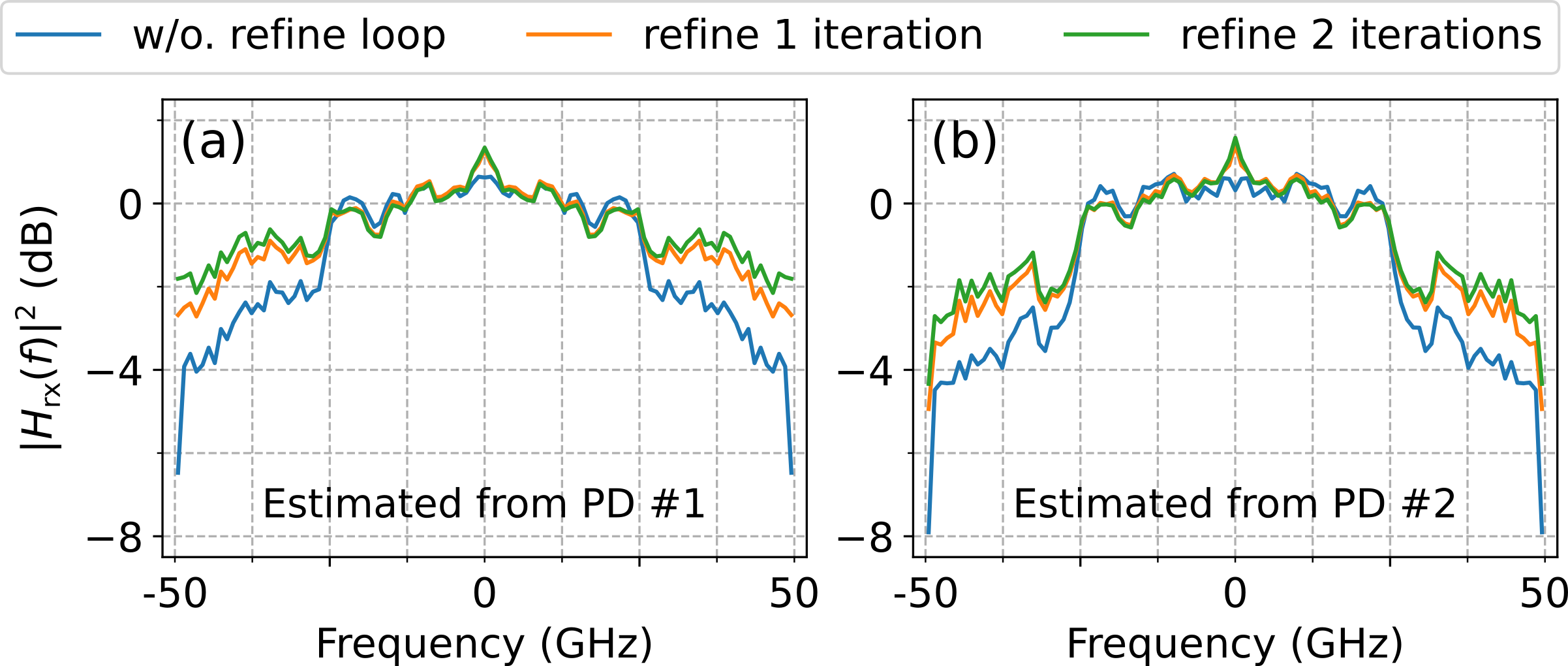}
    
	%\setlength{\abovecaptionskip}{-0.2cm}
	%\setlength{\belowcaptionskip}{-0.5cm}
	%\vspace{-3pt}
	%\vspace{-15pt}
	\centering\caption{Estimated receiver-side normalized power frequency responses obtained from (a) the first (dispersed branch) and (b) second photodiodes.}
	\label{fig5}
\end{figure}

Figures~\ref{fig5}(a,b) depict the Rx-side normalized power frequency response obtained from the two PDs with different refinement loop configurations.
These response results are estimated from the measured intensities of the training sequence after a 40-km transmission under a \mbox{35-dB} OSNR using the estimation method discussed in Subsection \Rmnum{4}-C.
The curves reveal that with increasing refinement iterations, the transfer characteristics for high-frequency components exceeding \mbox{$\pm$25~GHz} get improved.
Consequently, the 3-dB bandwidth of both ports converges towards \mbox{50~GHz}.

\begin{figure}[htbp]
	%\vspace{-12pt}
	\centering
	\captionsetup{justification=centering}
    \includegraphics[width=0.98\hsize]{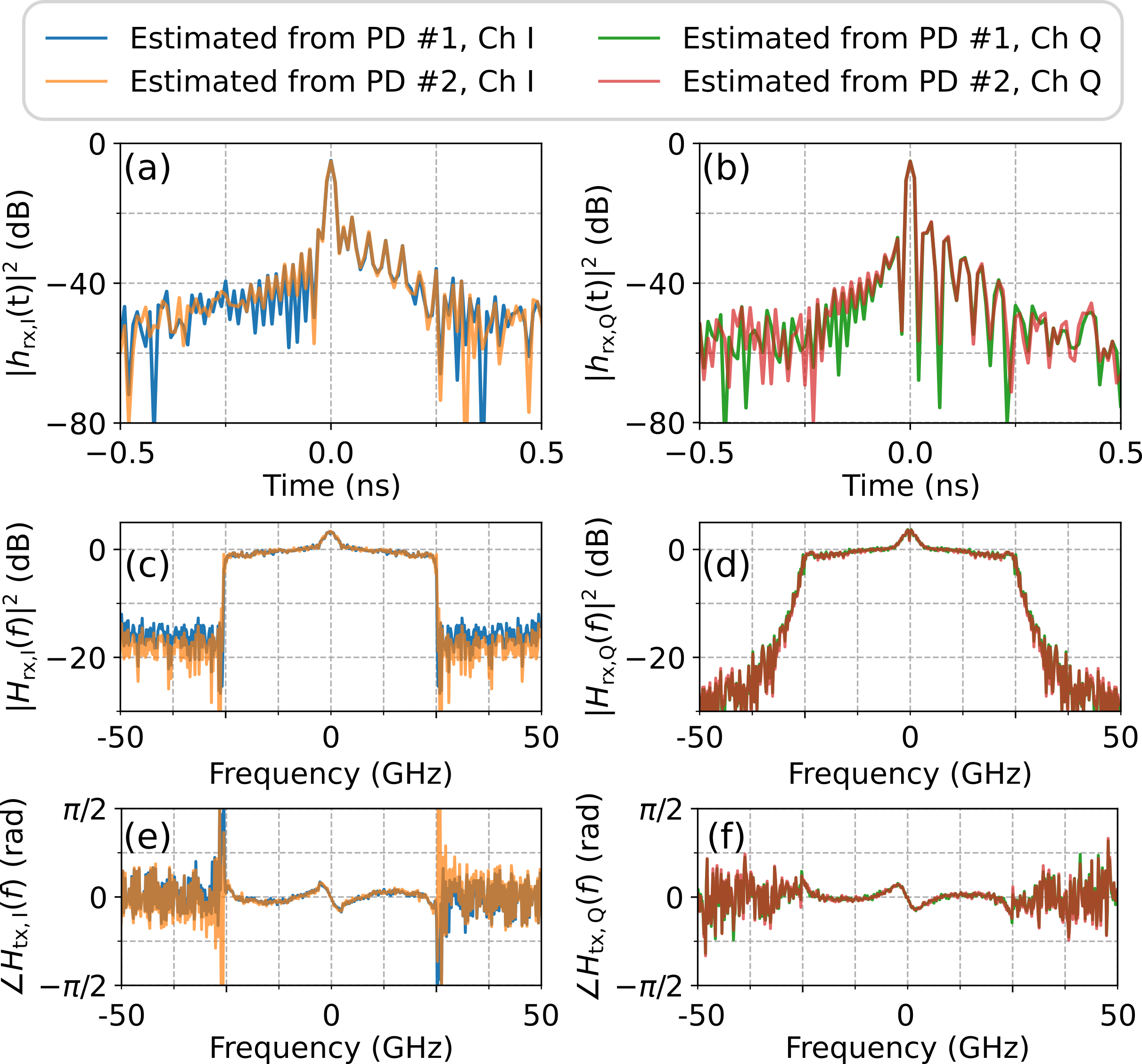}
    
	%\setlength{\abovecaptionskip}{-0.2cm}
	%\setlength{\belowcaptionskip}{-0.5cm}
	%\vspace{-3pt}
	%\vspace{-15pt}
	\centering\caption{Estimated transmitter-side (a,b) impulse power response, (c,d) normalized power and (e,f) phase frequency response of the I- and Q-channels.}
	\label{fig6}
\end{figure}

Figures~\ref{fig6}(a$\sim$f) present the Tx response results estimated from the same intensity data as in Figs.~\ref{fig5}(a,b) with 2 refined loops.
The I- and Q-channel are estimated separately since they possess different modulator nonlinear coefficients.
Figures~\ref{fig6}(a,b) show the power-normalized time domain impulse power response for the I and Q channels.
The power and frequency phase responses are shown in Figs.~\ref{fig6}(c,d) and Figs.~\ref{fig6}(e,f), respectively, which both exhibit structured features of modulation signals.
We notice that the results obtained from the two PDs appear to be similar in overall outline, implying that the effectiveness of the estimated method is likely to be robust to CD variation.

\begin{figure}[htbp]
	%\vspace{-12pt}
	\centering
	\captionsetup{justification=centering}
    \includegraphics[width=\hsize]{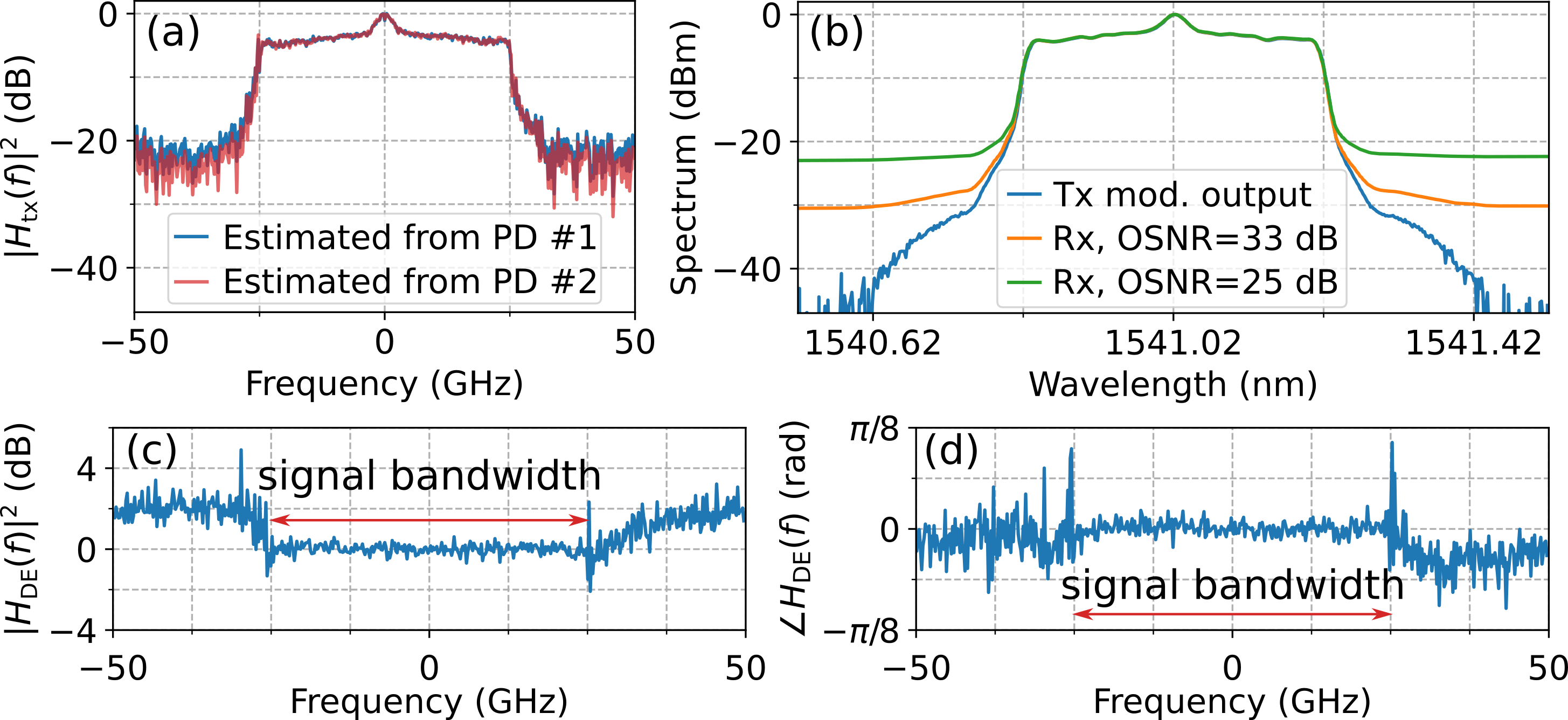}
    
	%\setlength{\abovecaptionskip}{-0.2cm}
	%\setlength{\belowcaptionskip}{-0.5cm}
	%\vspace{-3pt}
	%\vspace{-15pt}
	\centering\caption{(a) Estimated transmitter-side normalized power frequency response. (b) Measured optical spectra. Estimated (c) normalized power and (d) phase frequency response of the dispersive element.}
	\label{fig7}
\end{figure}
\begin{figure}[htbp]
	%\vspace{-12pt}
	\centering
	\captionsetup{justification=centering}
    \includegraphics[width=\hsize]{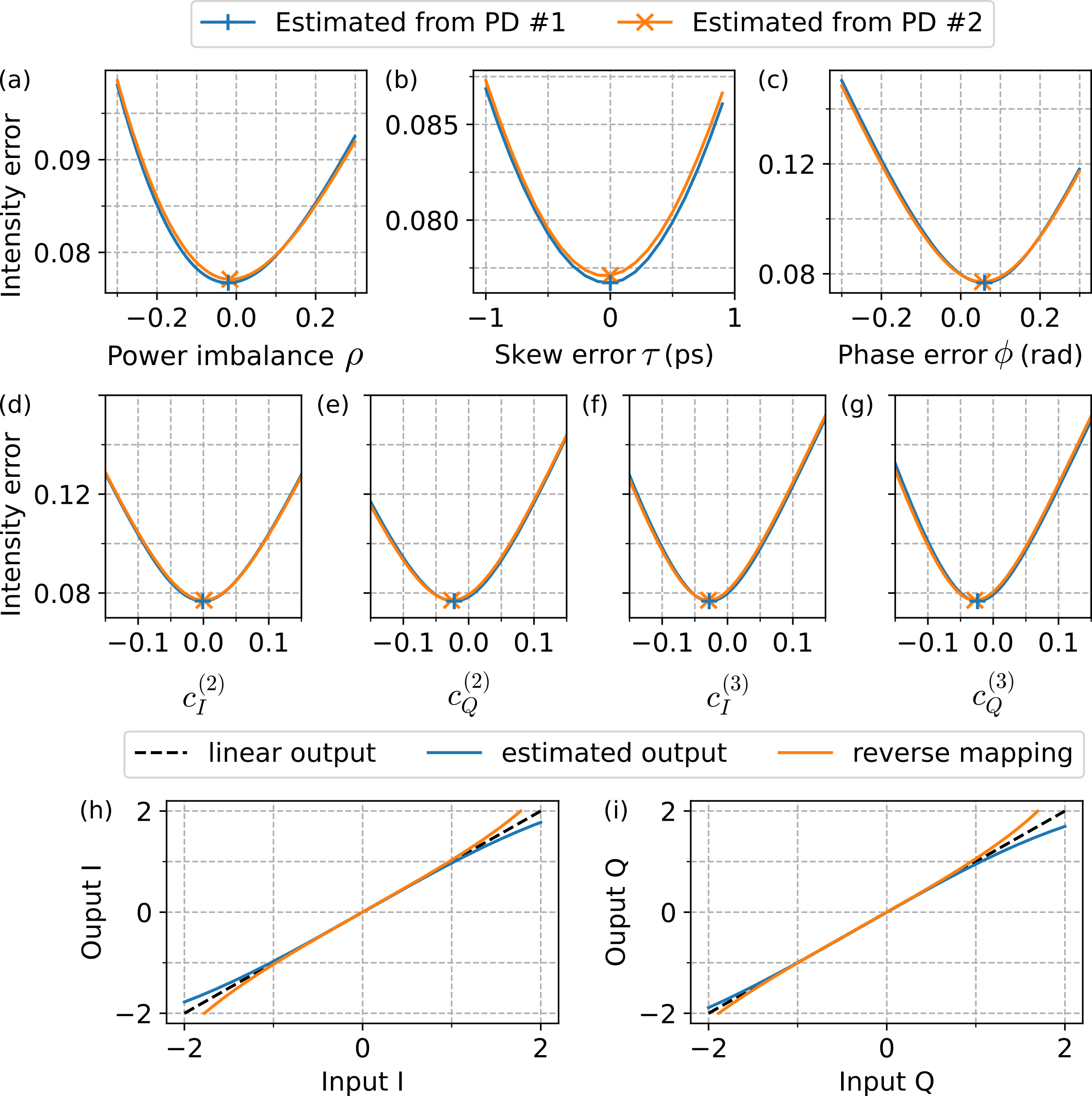}
    
	%\setlength{\abovecaptionskip}{-0.2cm}
	%\setlength{\belowcaptionskip}{-0.5cm}
	%\vspace{-3pt}
	%\vspace{-15pt}
	\centering\caption{(a$\sim$g) Mean absolute error curves as functions of parameters to be determined. Estimated modulator nonlinear transfer curve for (h) the I- and (i) Q-channels.}
	\label{fig8}
\end{figure}

After combining the estimated I- and Q-channel results, the Tx-side normalized power frequency response is shown in Fig.~\ref{fig7}(a).
The measured optical spectra using OSA at Tx and Rx with different OSNRs are given in Fig.~\ref{fig7}(b).
The spectrum humps within \mbox{$\pm$3~GHz} may arise due to the unflatness of the frequency transfer characteristics of the EAs or RF chains at the Tx, which can be clearly observed in both the measured optical and estimated electrical spectra.
By calculating the division of the estimated frequency response between the dispersed branch and the undispersed branch, Figs.~\ref{fig7}(c,d) present the estimated spectrum and frequency phase response of the dispersive element.
Although the small phase variation could be raised by the phase ripple of the FBG, the flatness of both the spectrum and phase profiles within the signal bandwidth suggests that the dispersive element introduces only negligible distortion.

Figures~\ref{fig8}(a$\sim$g) illustrate the resulting changing curves of intensity error as a function of each estimated parameter for $\phi$, $\tau$, $\rho$, $c_I^{(2)}$, $c_Q^{(2)}$, $c_I^{(3)}$, $c_Q^{(3)}$ during IQ-dependent impairment and modulator nonlinearity estimation.
The intensity error is determined by calculating the MAE between the measured and estimated intensities of the training sequence.
The estimated parameter values with the minimum intensity MAEs are marked with $+$ and $\times$ from two receiving photodiodes, respectively.
Although the two branches have different CD values, they obtain similar estimated values for these parameters.
Figures~\ref{fig8}(h$\sim$i) depict the modulator nonlinear transfer curve for the I- and Q-channels obtained by the estimated nonlinear coefficients from the dispersed branch.
The black dashed line represents the linear transfer function as a reference.
The orange curves denote the calculated nonlinear mapping to reverse the modulator nonlinearity.

\begin{figure}[htbp]
	%\vspace{-12pt}
	\centering
	\captionsetup{justification=centering}
    \includegraphics[width=0.95\hsize]{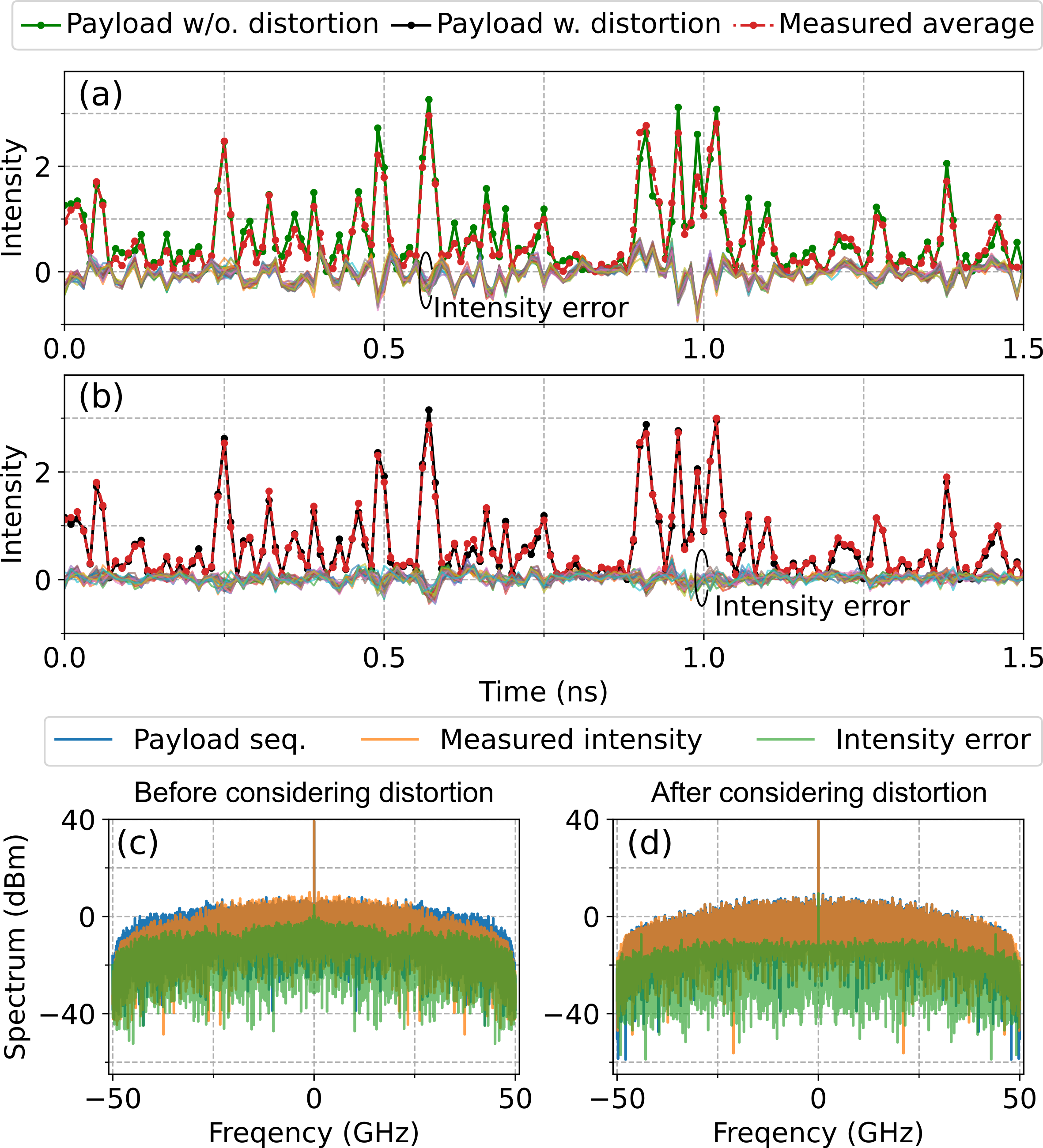}
    
	%\setlength{\abovecaptionskip}{-0.2cm}
	%\setlength{\belowcaptionskip}{-0.5cm}
	%\vspace{-3pt}
	%\vspace{-15pt}
	\centering\caption{Comparison between measured and estimated intensity waveforms (a)~without and (b)~with considering distortion induced by channel impairments, and (c,d)~their spectra in two cases.}
	\label{fig9}
\end{figure}

Figures~\ref{fig9}(a$\sim$d) validate the effectiveness of the channel estimation.
Figures~\ref{fig9}(a,b) compare approximation levels between the estimated and measured intensities for the undispersed branch, considering scenarios with and without estimated channel distortion.
In both figures, red dashed lines with dots depict the average intensity of 30 separate measurements.
The payload section of the transmitted sequence is assumed to be distorted by estimated impairments, including Tx response, IQ impairments, and modulator nonlinearity. 
Then, the sequence propagates to the Rx with estimated CD values.
Consequently, the expected intensity waveform can be deduced via the square-law detection formula, represented by a black line with dots in Fig.~\ref{fig9}(b).
In contrast, a green line with dots illustrates the expected intensity without considering the channel distortion in Fig.~\ref{fig9}(a).
The intensity errors for both schemes are calculated through 30 independently measured traces and depicted using lines with different colors.
The high OSNRs during measurements constrain the distribution of these intensity error waveforms.
Notably, numerous spikes emerge in the intensity error curves of Fig.~\ref{fig9}(a), while their counterparts in Fig.~\ref{fig9}(b) display minimal distinct spikes, a consequence of having knowledge regarding the channel distortion.
Figures~\ref{fig9}(c,d) contrast the errors between scenarios with and without the incorporation of estimated distortion, by presenting the spectra of both the estimated payload intensity and its measured counterpart. 
In both figures, the green shades denote the spectra of the intensity error using one measured trace.
The lower and flatter outline of the intensity error spectrum in Fig.~\ref{fig9}(d) indicates the error has been mitigated after propagating a distorted payload sequence.
The rest fluctuations in Fig.~\ref{fig9}(b) and the noise floor in Fig.~\ref{fig9}(d) could be attributed to the effective number of bits (ENOB) limitations of analog-to-digital converters (ADCs) and the residual estimation error during channel estimation.
The dispersed branch intensity appears similar approximation enhancement after applying channel distortion, while it is not shown by a figure to avoid repeated illustrations.
For both of the two branches, by propagating the payload sequence without channel distortions, the measured intensity and amplitude signal-to-noise ratios (SNRs) are around 14 and 16~dB.
After considering the estimated channel distortions, the intensity and amplitude SNRs for both branches achieve roughly 19 and $20$~dB, respectively.

\subsection{Reconstruction Results}

\begin{figure*}[t!]
	%\vspace{-12pt}
	\centering
	\captionsetup{justification=centering}
    \includegraphics[width=0.98\hsize]{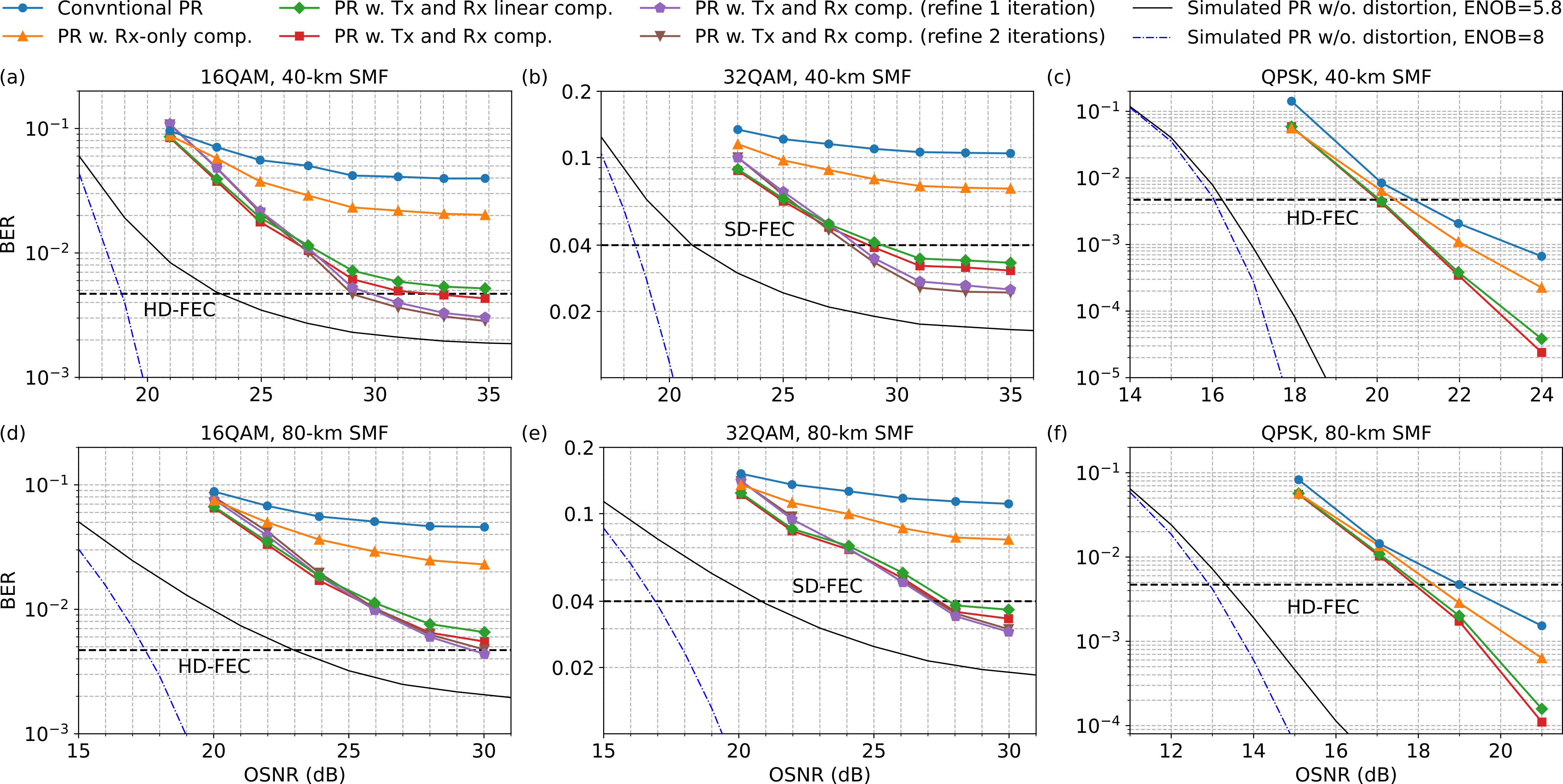}
    
	%\setlength{\abovecaptionskip}{-0.2cm}
	%\setlength{\belowcaptionskip}{-0.5cm}
	%\vspace{-3pt}
	%\vspace{-15pt}
	\centering\caption{Measured BER curves under various OSNRs with (a,d)~16QAM, (b,e)~32QAM, and (c,f)~QPSK signals over 40- and 80-km SMF transmission employing different PR schemes.
    Simulated BERs without channel-induced distortion are for reference.}
	\label{fig10}
\end{figure*}

\begin{figure*}[htbp]
	%\vspace{-12pt}
	\centering
	\captionsetup{justification=centering}
    \includegraphics[width=\hsize]{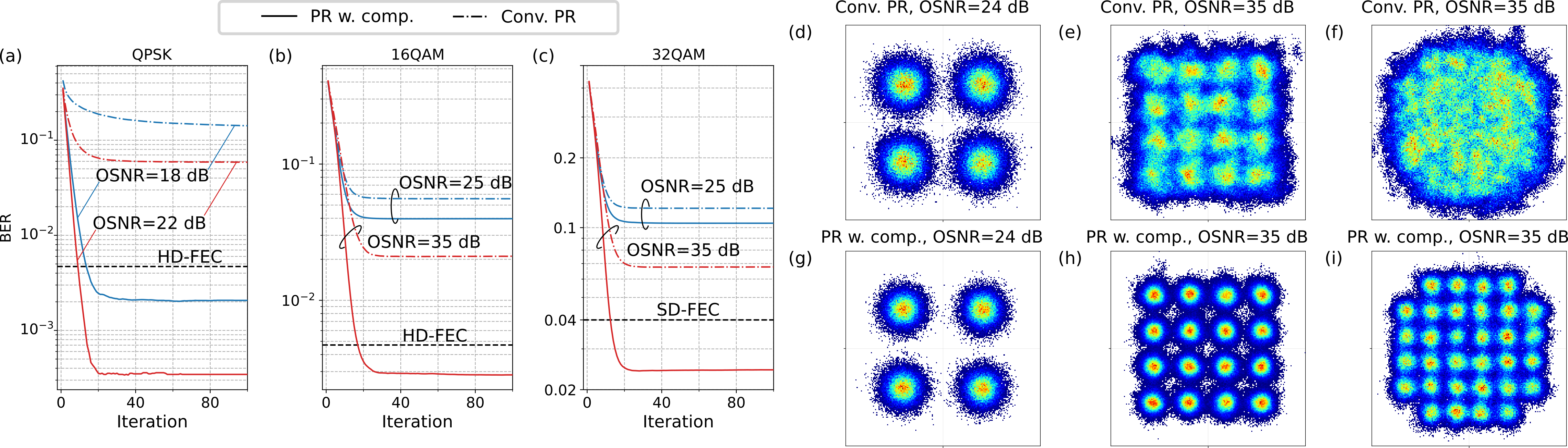}
    
	%\setlength{\abovecaptionskip}{-0.2cm}
	%\setlength{\belowcaptionskip}{-0.5cm}
	%\vspace{-3pt}
	%\vspace{-15pt}
	\centering\caption{Measured BERs versus reconstruction iteration number for (a)~QPSK, (b)~16QAM, and (c)~32QAM signals. Recovered constellations employing (d$\sim$f) the conventional PR and (g$\sim$i) proposed PR schemes.}
	\label{fig11}
\end{figure*}
Figures~\ref{fig10}(a$\sim$f) illustrate the measured pre-FEC BERs versus OSNRs for QPSK, 16QAM, and 32QAM signals across 40- and 80-km SSMF transmission with different reconstruction configurations.
In these figures, a saturated pilot symbol ratio of 50\% is adopted to explore the BER lower bounds, reflecting the maximum achievable reconstruction accuracy across different OSNRs.
In this case, since the BERs are constrained by noise rather than local minima, the selective phase reset operation is skipped.
More than $10^{5}$ symbols are used for the computation of each data point on the BER curves.
The distortion-aware PR with no refinement loop, following the procedure detailed in Subsection \Rmnum{2}-D, is delineated by red lines with squares.
The same PR schemes, but including 1 and 2 refinement loops, are depicted by purple and brown lines, respectively.
Blue lines denote the conventional PR that omits all distortion compensation techniques including FFE pre-equalization.
Orange lines with triangles denote the conventional PR that integrates FFE pre-equalization.
Green lines with diamonds represent the distortion-aware PR excluding compensation for IQ impairment and modulator nonlinearity.

In high OSNR regimes, e.g. exceeding 28~dB, the PR scheme employing all compensation techniques and refinement loops delivers the lowest BERs.
As OSNR falls below 28~dB, the distortion-aware PR with no refinement loop emerges as the optimal approach, suggesting that the estimation error tends to accumulate during refinement loops under noisy measurements.
Surprisingly, the 80-km transmission results reveals lower BERs in low OSNR regimes as compared to the 40-km transmission.
This deviation from conventional understanding can be attributed to the pump current difference of the Rx EDFA when measuring over two distances.
To avoid the power saturation issue at the highest OSNR, the 40-km transmission uses a pump current of 377~mA rather than 890~mA in the 80-km case, which leads to a received optical power decrease by a few decibels in low OSNR regime.
Therefore, the thermal noise at the Rx has more impact in 40-km transmission.
The simulated BER-OSNR curves, considering ASE, thermal noise, and ENOB, are given as a reference indicating the performance without any distortion from channel impairments.
The simulation assumes a PD responsivity of 1~A/W and a thermal noise level at \mbox{10e-12}~\mbox{A/Hz$^{1/2}$}.
The results with ENOBs of 5.8 and 8 are marked with black solid lines and blue dashed lines, respectively.
In the left two columns of the figures, the error floors of the simulated black BER curves indicate that the ENOB limitation could be the bottleneck hindering the high-order formats from having lower BERs.
%the influence of digital CD premixing and the methodology employed for inducing ASE within the experimental setup.
%While the ASE is introduced at the Rx, the estimated field needs to be propagated with the premixed CD during iterations.
%Consequently, this discrepancy means that the PR iteration does not propagate the ASE to the correct CD projections.
%The 80-km transmission results possess smaller CD values during the propagation of PR iterations due to the positive CD in fiber counteracts the negative CD from digital premixing, which may cause the ASE-induced errors to have a more mild impact on the reconstruction performance.

Figures~\ref{fig11}(a$\sim$c) depict the measured pre-FEC BER curves versus iteration number recovering QPSK, 16QAM, and 32QAM signals over a 40-km transmission under various \mbox{OSNRs} using the distortion-aware and conventional PR schemes.
Due to an abundant pilot symbol ratio of 50\%, both PR schemes achieve convergence within 40 iterations.
However, their converged BER floors appear to be different and are influenced by the PR configuration and the received OSNRs.
The conventional PR scheme exhibits high BER floors, failing to meet the FEC thresholds, which can be explained by the uncompensated distortion.
In contrast, the BER curves employing the distortion-aware PR satisfy FEC thresholds under relatively high OSNRs, while performing an earlier and degraded convergence under low OSNRs, validating the impact of noisy measurements on PR algorithms as analyzed in Subsection \mbox{\Rmnum{2}-A}.
The recovered constellations after convergence for both schemes are compared in Figs.~\ref{fig11}(d$\sim$i).
The compact clustering of the constellations demonstrates the effectiveness of the distortion-aware PR.

\subsection{Achievable Data Rate Assessment}

The previous subsection illustrates the achievable BER limits constrained by noise during measurements.
Yet, this does not offer a direct insight into the systematic achievable capacity of the presented experimental setup.
Given that pilot symbols occupy the time slots allocated for payload symbols, a trade-off exists between reconstruction accuracy and the pilot symbol ratio.
Consequently, the maximum data rate may not align with the highest reconstruction accuracy.

To determine the achievable data rates, we follow a similar procedure as described in~\cite{High_Capacity_Transmission_with_FMF}.
We implement encoding and decoding using low-density parity-check (LDPC) codes from the DVB-S2 standard~\cite{LDPC_DVB_S2} and an outer hard-decision code with 6.25\% overhead~\cite{6p25_HD_FEC}. 
The concatenated coding approach eliminates the issue of measuring a limited number of symbol in the experiments, allowing us to claim a valid data rate at post-LDPC BERs of below the outer FEC theshold rather than below $10^{-10}$.
However, embedding encoded patterns in the transmitter adds inconveniences to the experiments for a significantly increased measurement number and offers no assurance against time-varying factors (e.g. modulator bias drifting due temperature fluctuation) affecting the measured outcomes.
As a result, we choose to directly apply LDPC codes to the uncoded offline data by introducing a syndrome sequence and modifying the min-sum algorithm as described in~\cite{uncoded_FEC}.
The DVB-S2 LDPC codes consist of a set of different code rates with a fixed block length of 64,800 bits.
LDPC code rates of 1 (only employing the outer code), 9/10, 5/6, 3/4, 2/3, and pilot symbol ratios ranging from 1/2 to 1/6 are selected.
13 code blocks are utilized in computing the pre-LDPC BERs.
Due to a lack of variation in the transmitted pseudo-random pattern, we integrate a random interleaver placed after soft mapping to improve estimation precision.

\begin{figure}[htbp]
	%\vspace{-12pt}
	\centering
	\captionsetup{justification=centering}
    \includegraphics[width=0.98\hsize]{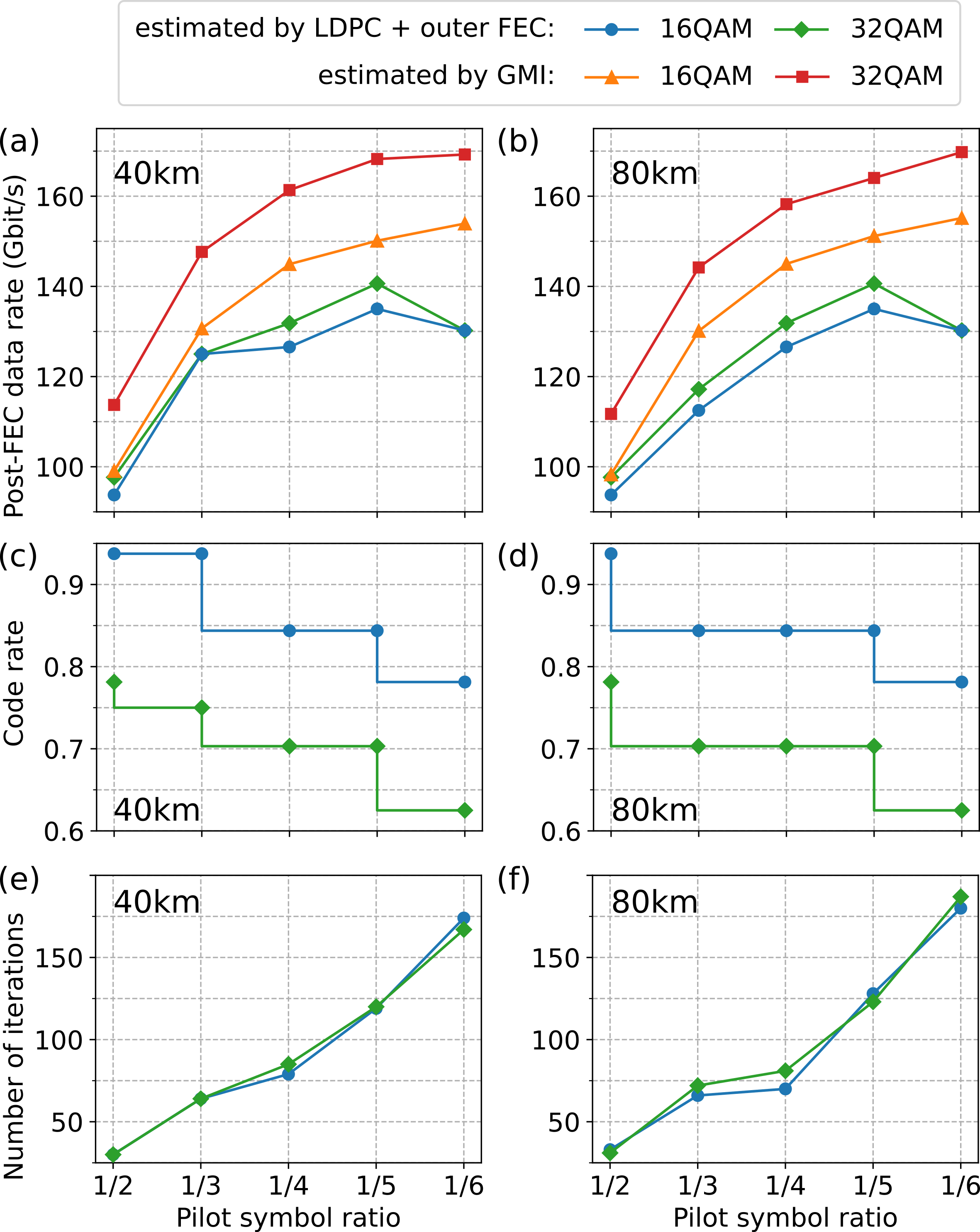}
    
	%\setlength{\abovecaptionskip}{-0.2cm}
	%\setlength{\belowcaptionskip}{-0.5cm}
	%\vspace{-3pt}
	%\vspace{-15pt}
	\centering\caption{Estimated post-FEC data rate versus different pilot symbol ratios for (a) 40- and (b) 80-km transmission. The corresponding (c,d) total code rates and (e,f) required numbers of iterations for the two distances.}
	\label{fig12}
\end{figure}

Figures~\ref{fig12}(a,b) show the resulting post-FEC net data rates versus different pilot symbol ratios for 16QAM and 32QAM signals over 40- and 80-km transmissions, evaluated at their peak OSNR values.
The selective phase reset operation is enabled, and the phase reset threshold is adjusted to its optimal value to drive the algorithm out of local minima and result the lowest pre-FEC BERs.
The net data rates are calculated by selecting the highest LDPC code rate that yields a post-LDPC BER below the outer FEC threshold ($4.7\times 10^{-3}$) for each pilot symbol ratio.
The GMI, reflecting the achievable data-rate upper limit employing an ideal soft-decision FEC code of infinite code length, is estimated based on the AWGN noise assumption and presented as a reference metric~\cite{GMI}.
The total code rates after considering the outer code are given in Figs.~\ref{fig12}(c,d) for the two transmission distances.
It can be seen that the maximum post-FEC data rates of 140~Gb/s are achieved at a pilot symbol ratio of 1/5 employing 32QAM modulation format for both distances, while the pilot symbol ratio of 1/2 only yields roughly 100~Gb/s.
When the pilot symbol ratio falls below 1/5, the LDPC-estimated net data rates start to decline due to the arising estimation error in the PR process.
Nonetheless, the estimated GMIs continue to increase, implying that the post-FEC data rates drop due to the limited granularity of the code rates and could be further enhanced using advanced modulation formats with enhanced rate adaptability, such as probabilistic constellation shaping (PCS).
%possibly because the estimation error deviates from the AWGN distribution, leading to more error bursts in the post-LDPC bit streams than GMIs suggest.
We also evaluated the conventional PR without considering channel distortions.
In this case, after 40-km transmission, the LDPC code rates obtaining maximum achievable post-FEC data rates are 3/4 and 1/2 at a pilot rate of 1/3 for 16QAM and 32QAM signals, respectively.
In contrast, the distortion-aware PR scheme demonstrates an approximate $\sim$49\% enhancement in the achievable post-FEC data rate compared with the conventional PR.
This can be ascribed to its improved reconstruction accuracy, thus reducing the dependency on the pilot symbols.
Figures~\ref{fig12}(e,f) depict the number of iterations required for the reconstruction algorithm to fully converge, which is determined by the relative change rate of mean amplitude error between the estimated fields of two traces in Fig.~\ref{fig4}.
A relative change rate of less than 0.1\% lasting for ten iterations is regarded as a sign of full convergence.
The computational complexity of PR predominantly arises from the (inverse) fast Fourier transform (FFT/IFFT) operations associated with complex-value field propagation in the reconstruction stage.
Thus, the computational complexity for each processing block can be estimated as $\mathcal{O}(KMN\log_2N)$, where $K$, $M$, $N$ represent the number of required iterations, the FFT/IFFT operations per iteration, and the block size, respectively.
The data in Figs.~\ref{fig12}(e,f) reveals that while high-order QAM formats do not significantly affect the required iteration count, increasing the pilot symbols tends to proportionally reduce $K$.
This suggests that under conditions of relatively low noise level, opting for an advanced QAM order choosing a high pilot symbol ratio might be more advantageous than simpler QAM formats with a low pilot symbol ratio to meet a specific data rate, since the former requires much fewer iterations to converge and thereby saves the computational complexity.

\subsection{Discussion}

To summarize the experimental findings, to begin with, the channel results obtained in the training stage offer useful information regarding the transmitter's frequency response and IQ-dependent impairments as well as modulator nonlinearity.
This suggests that a photodiode can function as a transmitter monitoring device without resorting to direct phase measurements.
Secondly, the proposed PR approach allows enhanced optical field reconstruction after considering distortion effects from various channel impairments.
The reconstruction process fully considers the propagation effect within the fiber, thus making the performance resilient to chromatic dispersion.

Regarding the performance limiting factors, in high OSNR regimes, if we assume that the thermal noise is negligible for adequate received optical power, then the reconstruction accuracy is constrained by the limited ENOB during analog-to-digital (A/D) conversion.
This limitation is fundamentally linked to the principle of PR, as symbol mixing inherently generates intensity waveforms with elevated PAPRs, resulting in a relatively low signal-to-quantization-noise ratio.
In the experiment, we observe PAPRs spanning from 13 to 17~dB in the captured intensity waveforms.
While clipping before A/D conversion can mitigate this effect, excessive clipping degrades the PR performance by distorting the intensity waveform peaks.
In the future, advancements in high-speed A/D technology may resolve the ENOB challenge.

The requirement on pilot symbol ratios is another issue.
As depicted in Figs.~\ref{fig12}(c,d), a reduction in pilot symbol ratios correlates with a decline in code rates and a proportional increase in the required iteration number.
This pattern underscores that the inherent illness of a two-PD-based PR Rx is not fully addressed without sufficient pilot symbol constraints.
Considering these factors, a carrierless PR approach that balances a moderate pilot symbol ratios with fast convergence and high precision might necessitate additional independent intensity measurements.
As the study on space-time diversity PR \cite{SpaceTime_PR} indicates, the BER floor issues at reduced pilot symbol ratio can be alleviated by employing four intensity measurements instead of two.
The contrast between \cite{SpaceTime_PR} and this work suggests two potential development paths for PR, both of which can benefit by utilizing the proposed distortion compensation strategy.
The first prioritizes the uses of numerous PDs, advanced modulation formats, and moderate pilot symbol ratios to maximize PR's transmission capacity.
The second leverages simper, noise-resilient modulation formats and a small number of PDs, such as two, to construct a simplified receiving front-end, and primarily employs PR as a transmission scheme that is both chromatic dispersion-resistant and colorless (resilient to wavelength drift and laser phase noise).

%-------------------------Section 6 ---------------------------%
\section{Conclusion}

In this paper, we investigated the feasibility and limitations of receiving high-order QAM signals using carrierless, intensity-only measurements through PR receiving techniques.
The errors observed in the received intensities do not follow the field propagation relationship across different measurement projections, thus limiting the converged accuracy of the PR reconstruction.
To mitigate part of these errors induced by channel impairments, a distortion-aware PR scheme has been proposed and demonstrated improved reconstruction accuracy in the experiments, which enabled the successful transmission over 40- and 80-km SSMF and achieved BERs below the 6.25\% HD-FEC and 25\% SD-FEC thresholds for 50-GBaud 16QAM and 32QAM signals, respectively.
By implementing concatenated coding schemes, the achievable post-FEC data rate reached 140~Gb/s per polarization per wavelength channel at an optimal pilot symbol ratio of 20\% for both distances.
Compared with carrier-assisted detection schemes, the PR approach requires no assistance from the local or signal-accompanying optical carrier.
As a result, PR receivers can benefit from having a simplified front-end architecture and enhanced energy efficiency in the optical domain, as it fully uses the transmitting optical power to send the information-carrying signals.
Therefore, PR could be a promising solution for high-speed short-reach optical interconnect scenarios, such as datacenter and metropolitan area networks.

% or


%\appendices

%\section{}

%\section*{Acknowledgment}

\ifCLASSOPTIONcaptionsoff
  \newpage
\fi



%\iffalse

%\fi

\bibliography{IEEEabrv,JLT_Remote_Pilots}

\end{document}